\documentclass[twocolumn]{aastex62}

\newcommand\kms{km\,s$^{-1}$}

\accepted{October 8, 2018}

\shorttitle{VLA measured SiO maser positions in evolved stars} 
\shortauthors{Pihlstr\"om et al.}
\begin{document}

\title{Positional Offsets Between SiO Masers in Evolved Stars and
  their Cross-Matched Counterparts}

\correspondingauthor{Ylva Pihlstr\"om}
\email{ylva@unm.edu}

\author[0000-0003-0615-1785]{Ylva M.~Pihlstr\"om}
\altaffiliation{Y.M.~Pihlstr\"om is also an Adjunct Astronomer at the\\
  National Radio Astronomy Observatory}
\affiliation{Dept. of Physics and Astronomy, University of New Mexico, 1919 Lomas Boulevard NE, Albuquerque, NM 87131, USA}

\author[0000-0003-3096-3062]{Lor\'ant O.~Sjouwerman}
\affiliation{National Radio Astronomy Observatory, 1003 Lopezville
  Road, Socorro, NM 87801, USA}

\author{Mark J~Claussen}
\affiliation{National Radio Astronomy Observatory, 1003 Lopezville Road, Socorro, NM 87801, USA}

\author[0000-0002-6753-2066]{Mark R.~Morris}
\affiliation{Department of Physics and Astronomy, University of California, Los Angeles, CA 90095-1547, USA}

\author[0000-0003-0427-8387]{R.~Michael Rich}
\affiliation{Department of Physics and Astronomy, University of California, Los Angeles, CA 90095-1547, USA}

\author[0000-0002-0230-5946]{Huib Jan van Langevelde}
\affiliation{Joint Institute for VLBI ERIC, Postbus 2, 7990 AA Dwingeloo, The Netherlands}
\affiliation{Leiden University, Postbus 9513, 2300 RA Leiden, The Netherlands}

\author[0000-0002-9390-955X]{Luis Henry Quiroga-Nu{\~n}ez}
\affiliation{Joint Institute for VLBI ERIC, Postbus 2, 7990 AA Dwingeloo, The Netherlands}
\affiliation{Leiden University, Postbus 9513, 2300 RA Leiden, The Netherlands}

\begin{abstract}
  Observations of dust-enshrouded evolved stars selected from infrared
  catalogs requiring high positional accuracy, like infrared
  spectroscopy or long baseline radio interferometric observations,
  often require preparational observational steps determining a
  position with an accuracy much better than 1\arcsec. Using
  phase-referencing observations with the Very Large Array at its
  highest resolution, we have compared the positions of SiO 43 GHz
  masers in evolved stars, assumed to originate in their infrared
  detected circumstellar shells, with the positions listed in the {\it
    MSX}, {\it WISE}, 2MASS, and {\it Gaia} catalogs. Starting from an {\it
    MSX} position it is, in general, simple to match 2MASS and {\it WISE}
  counterparts. However, in order to obtain a {\it Gaia} match to the
  {\it MSX} source it is required to use a 2-step approach due to the
  large number of nearby candidates and low initial positional
  accuracy of the {\it MSX} data.  We show that the closest comparable
  position to the SiO maser in our limited sample \emph{never} is the
  {\it MSX} position. When a plausible source with a characteristic
  signature of an evolved star with a circumstellar shell can be found
  in the area, the best indicator of the maser position is provided by
  the {\it Gaia} position, with the 2MASS position being second-best.
  Typical positional offsets from all catalogs to the SiO masers are
  reported.

%% 250 word limit

\end{abstract}

\section{Introduction} \label{sec:intro} The Bulge Asymmetries and
Dynamical Evolution (BAaDE) project is surveying more than 28,000
color-selected red giant stars in the Galactic plane for SiO maser
emission (L.O.\,Sjouwerman et al., in prep.).  With an
instantaneous detection rate well over 50\%, a unique sample of
dynamical tracers in the plane is being constructed. At the
frequencies of the SiO maser (43 GHz and 86 GHz) visual extinction is
not a hinder, and extremely accurate line-of-sight stellar velocities
($\lesssim$ 2 \kms) are determined at the locations of the stars  \citep[][and references therein]{habing96}.  The number of sources
will be large enough to trace complex kinematic structures and
minority populations.  The velocity structure of these tracers will be
compared with the kinematic structures seen in molecular gas and other
objects near the Galactic Center, and thereby highlight kinematically
coherent stellar systems, complex orbit structure in the bar, or
stellar streams resulting from recently infallen
systems. Investigations of the bar and bulge dynamics have begun using
a subset of this new kinematic information in the inner Galaxy region
\citep{trapp18}.

The BAaDE survey also identifies sufficiently luminous SiO masers
suitable for follow-up parallax and proper motion determination using
very long baseline interferometry (VLBI).  With VLBI, it may be
possible to investigate in detail orbits of stars constituting the
stellar bar. Spectroscopic infrared (IR) data of the targets will also
be taken to investigate metallicity effects across the bar and bulge
region. Such follow-up studies require positional accuracies of the
targets of the order of 0\farcs 1 or less. As the general BAaDE
observing strategy using the NSF's Karl G.\ Jansky Very Large Array
(VLA) in the C and D configurations with resolutions of 1-2\arcsec\
utilizes the masers themselves for phase corrections (L.O.\,Sjouwerman et
al., in prep.), the known positional accuracy is not improved
beyond that of the initial {\it Midcourse Space Experiment} ({\it
  MSX}) positions \citep[1-2\arcsec;][]{egan03}.  To improve the SiO
maser target positions, proper VLA A-array phase-referencing
observations could be performed instead, pushing the accuracy down
below 0\farcs 01, e.g.\ as shown here. However, doing such
observations is impractical for several reasons. First, there are very
few suitable VLA 43 GHz calibrators in the plane, severely limiting
the number of sources that could be observed in this fashion.  Second,
phase referencing is time consuming, and re-observing the detected
sample in this mode would in principle mean tripling our original time
request at the telescope.  Alternatively, we investigate whether
cross-matching the parent {\it MSX} positions with other general
all-sky IR and optical catalogs, like the Two-Micron All Sky Survey
\citep[2MASS;][]{skruts06}, the Wide-field Infrared Survey Explorer
\citep[{\it WISE};][]{wright10}, and the {\it Gaia} Data Release 2
\citep[DR2;][] {gaiacoll16, gaiacoll18} with typical claimed absolute
positional accuracies $\lesssim$ 0\farcs 1, can be used to improve the
positional information.  If so, the intermediate phase-referencing
observations with the VLA in extended configurations may not be
required to obtain sufficiently accurate positions for follow-up
studies.

We here report on a limited study using the VLA in a regular
phase-referencing observing scheme to determine the positions of a set
of masers to $<$0\farcs 01 accuracy. The resulting positions of the
masers are compared to matched {\it MSX}, {\it WISE}, 2MASS and {\it Gaia}
positions, in order to obtain a limited empirical determination of the
positional agreement between the IR/optical and radio data.  While some
other catalogs with claimed accurate astrometry exist, they were not
included in our study due to their more limited sky coverage.

\section{Data collection} \label{sec:datacoll}

\subsection{Source Selection}\label{sec:selec}
Phase-referenced observations at the VLA were used to achieve accurate
positions of the SiO maser. The accuracy of the derived positions
depends in part on the goodness of the calibrated phases, requiring a
bright calibrator source with good positional accuracy, located near
the target field.  The VLA calibrator J1755$-$2232 is positioned in
the inner Galactic plane and has a listed brightness of 0.32~Jy/beam
in the VLA calibrator manual along with an absolute positional
accuracy quoted between 0\farcs 002$-$0\farcs 01, and was therefore
chosen to be the phase-referencing calibrator. Within a distance of
0\fdg 6 of J1755$-$2232, 33 previously BAaDE detected SiO maser stars
were observed in this experiment. The initial field centers used for
observing these targets were the {\it MSX} positions included in Table
\ref{tab:pos}.

\subsection{VLA Observations and Data Reduction}\label{sec:obs}
The sources were observed in June 2015\footnote{This date, 2015.5,
  coincides with the epoch of {\it Gaia} DR2.} under project code
15A-497 with the VLA in the A-array configuration, yielding an angular
resolution of about 50-100 milli-arcsecond (mas). The Doppler-shifted
frequencies of both the $^{28}$SiO(1$-$0) $v=2$ and $v=1$ lines were
covered by our setup (600 \kms~total velocity bandwidth). A
phase-referencing cycle time of 50 seconds was used, with 20 seconds
on the calibrator bracketing each target source, which in turn was
observed for 30 seconds. Each target was observed twice, resulting in
a typical 1.7 \kms\ (250 kHz) channel rms noise of 13 mJy/beam,
agreeing with the estimated theoretical rms noise for observations at
low elevations and 1 minute on-source integration.

The data were calibrated using the AIPS package, and a deconvolved map
of each maser was constructed using the CLEAN algorithm with a robust
weighting of zero.  The resulting synthesized beam sizes were
almost identical, 0\arcsec.128$\times$0\arcsec.045 at a position angle
of 30$^\circ$.  Of the 33 targets 26 were detected. Because SiO masers
are variable throughout the stellar cycle with a period of a few
hundred days, this is a likely reason for the seven non-detections.
For the detections, a two-dimensional Gaussian fit was performed to
determine the peak flux position of the emission, listed as the VLA
positions in Table \ref{tab:pos}, determined from the channel with
peak emission of the $v=1$ or $v=2$ lines.  On average the
signal-to-noise ratio (S/N) of the detection in one channel was 25,
leading to 1-$\sigma$ uncertainties in the reported SiO positions of
about 1.0 mas in $x$ and 2.8 mas in $y$ using the relation $\Delta
\theta_i = (0.54\,\theta_{i})/(S/N)NR$, where $\theta_{i}$ is the
full-width at half maximum of the synthesized beam in the
$i$-direction \citep{reid88}.  Given the beam position angle, the
resulting errors in Right Ascension (R.A.) and declination (decl.) are
approximately 0.8 and 2.4 mas. Spectra of these sources, along
  with the line properties, will be published as part of the main
  BAaDE project; see L.O.\,Sjouwerman et al.\ (in prep.).

\subsection{\emph{MSX}, {\it WISE} and 2MASS Data}\label{sec:msxdata}

%\startlongtable
%\begin{rotatetable*}
\begin{deluxetable*}{cccccclccc}
\tablecaption{Source VLA and IR/optical catalog positions\label{tab:pos}}
\tablehead{
\multicolumn{2}{c}{VLA} & \multicolumn{2}{c}{\it MSX} & \multicolumn{2}{c}{{\it WISE}} & \multicolumn{2}{c}{2MASS} & \multicolumn{2}{c}{\it Gaia}\\
\colhead{R.A.\,(h:m:s)} & \colhead{decl.\, ($^\circ$:\arcmin:\arcsec)} & \colhead{R.A.\, (m:s)} & \colhead{decl.\, (\arcmin:\arcsec)} & \colhead{R.A.\, (m:s)} & \colhead{decl.\, (\arcmin:\arcsec)} & \colhead{R.A.\, (m:s)} & \colhead{decl.\, (\arcmin:\arcsec)} & \colhead{R.A.\, (m:s)} & \colhead{decl.\, (\arcmin:\arcsec)}}
\startdata
17:57:45.7491&$-$22:40:37.634&57:45.74&40:37.9&57:45.748&40:37.60&57:45.747&40:37.62&57:45.7487&40:37.635\\
17:56:35.1030&$-$22:38:16.087&56:35.09&38:16.1&56:35.118&38:15.92&56:35.110&38:15.91&56:35.1038&38:16.066\\
17:57:16.8026&$-$22:37:19.622&57:16.78&37:21.0&57:16.802&37:19.47&57:16.810&37:19.56&57:16.8023&37:19.631\\
17:57:37.7290&$-$22:37:07.215&57:37.73&37:08.0&57:37.738&37:07.15&57:37.730&37:07.27&57:37.7296&37:07.201\\
17:57:32.9523&$-$22:33:21.200&57:32.93&33:22.3&57:32.955&33:21.18&57:32.961&33:21.03&57:32.9523&33:21.200\\
17:57:32.1456&$-$22:30:19.551&57:32.09&30:20.5&57:32.139&30:19.49&57:32.147&30:19.58&57:32.1452&30:19.555\\
17:55:19.7673&$-$22:28:49.418&55:19.82&28:49.1&55:19.757&28:49.44&55:19.773&28:49.43&55:19.7672&28:49.415\\
17:53:17.0344&$-$22:26:02.737&53:17.06&26:02.4&53:17.034&26:02.61&53:17.040&26:02.70&53:17.0352&26:2.7303\\
17:57:33.5178&$-$22:24:26.355&57:33.50&24:28.1&57:33.543&24:26.07&57:33.522&24:26.19&57:33.5195&24:26.340\\
17:54:53.7339&$-$22:22:30.810&54:53.76&22:30.4&54:53.750&22:31.00&54:53.740&22:30.60&54:53.7356&22:30.809\\
17:53:29.5411&$-$22:21:46.851&53:29.59&21:47.2&53:29.538&21:46.82&53:29.542&21:46.77&53:29.5411&21:46.854\\
17:56:48.5306&$-$22:17:41.335&56:48.50&17:42.4&56:48.552&17:41.29&56:48.542&17:41.35&56:48.5303&17:41.347\\
17:57:16.5811&$-$22:15:20.805&57:16.56&15:21.6&57:16.594&15:20.90&57:16.590&15:20.69&-         &-\\
17:56:22.7935&$-$22:13:48.308&56:22.78&13:50.2&56:22.794&13:47.76&56:22.792&13:48.15&56:22.7939&13:48.312\\
17:54:52.1393&$-$22:11:00.323&54:52.15&11:01.0&54:52.140&11:00.31&54:52.147&11:0.319&54:52.1394&11:0.3669\\
17:55:04.2674&$-$23:11:22.130&55:04.20&11:21.8&55:04.271&11:22.07&55:04.261\tablenotemark{a}&11:22.03&55:04.2651&11:22.159\\
17:54:10.0437&$-$23:06:36.299&54:10.06&06:35.6&54:10.039&06:36.13&54:10.047&06:36.25&54:10.0436&06:36.289\\
17:56:11.9651&$-$23:04:28.685&56:12.02&04:27.8&56:11.978&04:28.06&56:11.959&04:28.70&56:11.9648&04:28.683\\
17:54:16.2127&$-$23:02:35.897&54:16.20&02:35.2&54:16.216&02:35.80&54:16.212&02:35.98&54:16.2126&02:35.885\\
17:55:05.1692&$-$23:01:42.724&55:05.21&01:41.9&55:05.169&01:42.54&55:05.173\tablenotemark{a}&01:42.54&55:05.1679&01:42.730\\
17:54:16.7405&$-$23:01:36.955&54:16.75&01:36.5&54:16.748&01:36.93&54:16.742&01:36.98&54:16.7398&01:36.947\\
17:53:56.1170&$-$23:00:23.772&53:56.06&00:23.4&53:56.128&00:24.95&53:56.112&00:23.70&53:56.1170&00:23.763\\
17:53:30.2574&$-$22:55:30.018&53:30.22&55:29.3&53:30.252&55:29.85&53:30.253\tablenotemark{a}&55:29.76&53:30.2569&55:30.009\\ 
17:57:06.7550&$-$22:44:54.598&57:06.77&44:54.6&57:06.740&44:54.40&57:06.764&44:54.47&57:06.7552&44:54.596\\
17:55:32.0723&$-$22:05:07.412&55:32.06&05:08.5&55:32.060&05:07.62&55:32.068&05:07.43&55:32.0726&05:07.416\\
17:54:11.9268&$-$22:03:57.408&54:11.95&03:57.6&54:11.960&03:57.53&54:11.930&03:57.42&54:11.9274&03:57.407\\
\enddata
\tablenotemark{a}{Two candidate matches were found within the 5\arcsec\
  search radius. The selected cross-match is the reddest, brightest
  and also the closest candidate.}
\end{deluxetable*}
%\end{rotatetable*}

The BAaDE sources were originally selected from the {\it MSX} Point Source
Catalog version 2.3 \citep{egan03}, based on their IR color in order
to optimize for the detection of SiO masers \citep{sjouwerman09}.  The
{\it MSX} mission was designed to collect IR photometry along the full
Galactic plane, and in regions not covered by {\it IRAS}.  For regions
toward the Galactic center in particular, where {\it IRAS} was heavily
confused due to the high source density, {\it MSX} significantly
improved existing IR catalogs.  {\it MSX} has a beam of 18\farcs 3,
with astronomically useful bands observed at 8.3, 12.1, 14.7 and 21.3
$\mu$m (bands {\it A}, {\it C}, {\it D}, and {\it E}, respectively).
The {\it MSX} positional information is dominated by information from
band {\it A} as it is the most sensitive band along with having the
shortest wavelength.  The {\it MSX} astrometric accuracy depends on
the detection quality \citep{egan03}, and the sources selected for the
BAaDE sample used quality photometry flags $Q_X\geq 3$ (i.e., 3 and 4)
where $Q_X$ indicates the quality in band $X$, translating into a
positional accuracy between 0\farcs 80--1\farcs 7.

Based on the {\it MSX} positions and their accuracy, for each {\it
  MSX} position with a maser detected, a search radius
of 5\arcsec\ was used to search the NASA/IPAC Infrared Science
Archive (IRSA) for cross-matches in the {\it WISE} and 2MASS surveys, with
the results listed in Table \ref{tab:pos}.  

The {\it WISE} survey scanned the sky at 3.6, 4.6, 12 and 22 $\mu$m (bands
W1, W2, W3, and W4, respectively) with angular resolutions of 6\farcs
1, 6\farcs 4, 6\farcs 5 and 12\farcs 0.  Initially, the {\it WISE} All-Sky
catalog positions were referenced with respect to the 2MASS catalog,
giving accuracies of $\approx$0\farcs 15, but the more recent All{\it WISE}
catalog has since improved upon the accuracy to $<$0\farcs 1 by also
implementing proper motion for reference objects.  Within the search
radius only single cross-matches were found, with reported {\it WISE}
positional accuracies of on average 0\farcs 035 and 0\farcs 036 in R.A.\,
and decl.

% http://wise2.ipac.caltech.edu/docs/release/allwise/expsup/sec2_5.html

The 2MASS project observed the full sky with a resolution of 2\arcsec,
using the 1.24, 1.66 and 2.16$\mu$m ($J$, $H$, and $K_s$) bands
\citep{skruts06}.  Cross-matches to all of our SiO maser detected {\it MSX} sources were found in
the 2MASS Point Source Catalog, with three fields showing two
possible cross-matches within the search radius (marked in Table
\ref{tab:pos}).  Given the anticipation of our targets being
dust-enshrouded evolved stars, most likely asymptotic giant branch
(AGB) stars, the source which was reddest and brightest was selected,
which in all three cases also corresponded to the closest match in
position.  The quoted accuracies for all 2MASS positions are 0\farcs
06 in both R.A.\, and decl.

\subsection{\emph{Gaia} Data}\label{sec:gaiadata}
The {\it Gaia} mission is conducting a full sky survey at 0.7$\mu$m
($G$-band). Although the spectral energy distribution (SED) from
dust-enshrouded stars has its peak in the (near-)IR, the specific
selection for stars with thinner shell envelopes in the BAaDE survey
(i.e., Miras instead of OH/IR stars) allows on occasion for optical
emission being detected, for example, by {\it Gaia}. Whereas we do not
anticipate many {\it Gaia} counterpart matches in the most obscured
regions in the Galactic plane and for the thicker shell objects in the
entire BAaDE sample, the
sensitive {\it Gaia} data is expected to yield some counterpart
matches that can be studied here along with the IR catalogs.

The {\it Gaia} DR2 provides high quality astrometric data
\citep{gaiacoll18,lindegren18}, which was used to search for
cross-matches to our SiO maser sample.  The {\it Gaia} data, however, had to be
treated differently than the IR data, as searching within 5 arcseconds
of the {\it MSX} positions provides up to 8 matches for individual {\it MSX}
sources.  Given the uncertainty in the {\it MSX} positions, a smaller
search radius could not be applied without the risk of missing the
correct cross-match.  To ensure the correct candidate was selected,
color and brightness criteria were applied, similar to what was done
for the 2MASS multiple candidates (Sect.\ \ref{sec:msxdata}).  {\it
  Gaia} DR2 contains photometry for the full $G$ band covering
0.33-1.05$\mu$m, and for some sources also the photometry and associated
color measured with the integrated $G_{BP}$ and $G_{RP}$ bands at
0.33-0.68$\mu$m and 0.63-1.05$\mu$m, respectively \citep{evans18}. As the
targeted objects typically are large-amplitude variable stars, another
characteristic that could be used in ensuring a proper cross-match is
a variability indication. Only one candidate counterpart with this
information was found in the {\it Gaia} data for our sample, and thus
we ignored this characteristic further in our matching
scheme\footnote{For calculating a measure of variability other than
  using the {\it Gaia} DR2 variability flag, see e.g.\
  \citet{quiroga18,belokurov17}. This is beyond the scope of this
  paper.}.  In addition, the positional offsets between the {\it
  Gaia}-selected candidates and the VLA SiO masers, as well as the
offsets between the {\it Gaia} candidates and the previously
cross-matched 2MASS positions were considered to aid in the {\it Gaia}
cross-matching procedure:

\begin{itemize}
\item For the set of 26 masers, one source had no {\it Gaia} candidate
  cross-match at at all, possibly due to optical extinction.  Note
  that this region around the calibrator is at G006.63+1.38, a region
  for which not much extinction is expected, which may explain the
  large fraction of optical counterparts. The remaining 25 sources had
  a combined number of 78 candidate cross-matches in the {\it Gaia}
  DR2 catalog within a 5\arcsec\ radius of the {\it MSX} position. The
  {\it Gaia} $G$ band magnitude was collected for all 78 candidates,
  as was the $G_{BP}-G_{RP}$ color, which existed for 47 candidates.

\item A cross-match was determined to be the most likely match if the
  $G_{BP}-G_{RP}$ color (if existing) was the reddest
  amongst the candidates, and if the  positional offset to the SiO maser was the smallest and $<0$\farcs
  5.  For 14 of the 25 sources with {\it Gaia}
  data, such cross-matches existed. It turns out that for all these 14 sources,
  also the {\it Gaia}-2MASS offset was consistently the smallest and
  $<0$\farcs 5.

\item Subsequently, returning to the original 78 candidates and
  selecting on {\it Gaia}-2MASS offsets only, the same 14 candidates
  with {\it Gaia} colors were selected along with 11 additional
  cross-matches for those lacking color information\footnote{This
    implies that a 2-step matching scheme, following the sequence {\it
      MSX}$\rightarrow$2MASS$\rightarrow${\it Gaia}, is the best
    approach also when no accurate phase-referenced SiO maser positions
    are available as is the case for the majority of the BAaDE sample.}.
\end{itemize}

\begin{figure}[t]
\begin{center}
\includegraphics[width=8.5cm]{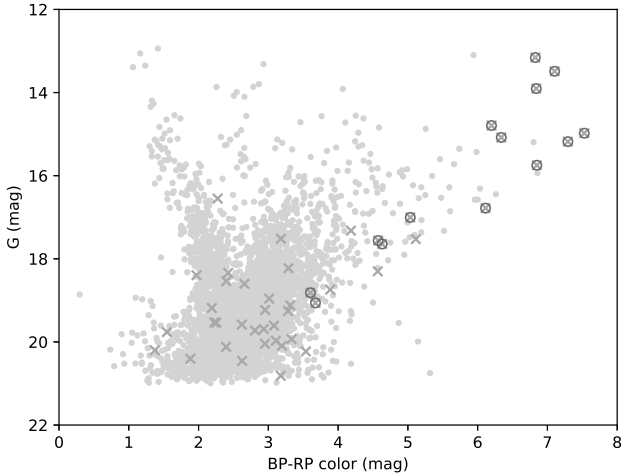}
\caption{A {\it Gaia} color-magnitude diagram, with 5,758 randomly
  selected sources in the neighborhood of the calibrator and target
  SiO masers plotted as light gray dots.  All the 47 {\it Gaia}
  candidate matches for which {\it Gaia} colors were available are
  marked by crosses, showing the spread in the diagram of all
  candidates.  After applying an angular distance offset and a color criterion,
  the 14 circles denote the selected cross-matches. This illustrates
  that the applied selection methodology primarily chooses redder and
  brighter stars, consistent with our targets being mainly redder
  dust-enshrouded (AGB)
  stars.\label{gaiacmd}}
\end{center}
\end{figure}

Figure \ref{gaiacmd} presents a color-magnitude diagram showing the
distribution of the 47 candidate cross-matches within 5\arcsec\
(crosses) on top of a set of 5,758 randomly selected {\it Gaia}
sources in the neighborhood of our targets, illustrating the spread of
the colors of the candidates. We note that, assuming an $M_{\rm bol}
\approx -6$ for the brightest AGB stars, a {\it Gaia} magnitude
fainter (larger) than 12 for this sample (Fig.\ 1) also indicates that
our counterparts are more distant than 4 kpc. We therefore may ignore
any {\it Gaia} parallax (i.e., $\pi < 0.25$ mas) and proper
motion\footnote{See first footnote.}  corrections here. The 14 circles
denote the position in the diagram of the candidates with reddest
color and offsets $<$0\farcs 5 from the SiO masers.  This demonstrates
that the selected cross-matches belong to the overall redder and
brighter population of stars, consistent with them being redder
dust-enshrouded (AGB) stars. The linearly averaged reported accuracies
for the {\it Gaia} positions of these faint sources were, on average,
0.99 mas (R.A.\,) and 0.64 mas (decl.).

\begin{figure}[t]
\begin{center}
\includegraphics[width=8.5cm]{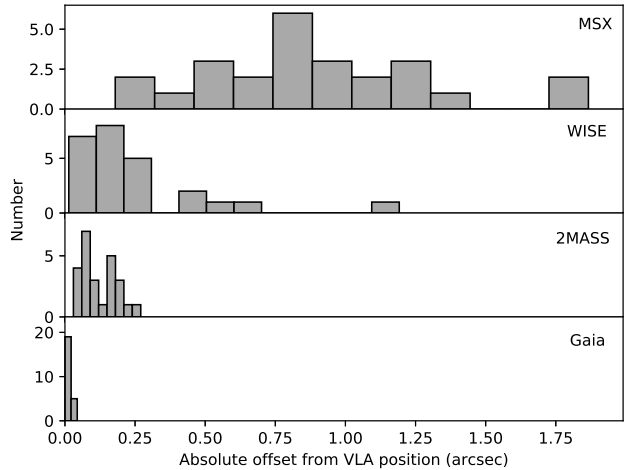}
\caption{Measured absolute offsets in arcseconds between the VLA SiO
  maser positions and the {\it MSX}, 2MASS, {\it WISE} and {\it Gaia} positions.
  The closest positional match is provided by {\it
    Gaia} if available, and by 2MASS otherwise.\label{offsets}}
\end{center}
\end{figure}

\begin{figure*}[t]
\begin{center}
\includegraphics[width=14cm]{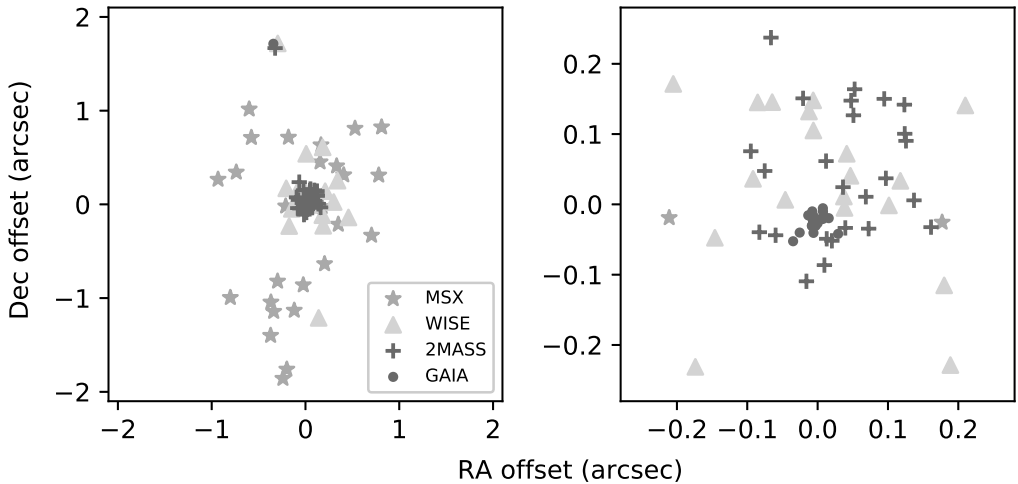}
\caption{Diagram of measured offsets in arcseconds between
  the VLA SiO maser positions and the {\it MSX}, {\it WISE}, 2MASS and {\it
    Gaia} positions.  The panel on the left hand side shows the full
  distribution (which is located well within the 5\arcsec\ search radius), while the right hand side is a zoom of the central
  region to emphasize the even closer match of the {\it Gaia} versus
  2MASS positions.\label{radecscatter}}
\end{center}
\end{figure*}

\section{Results} \label{results} 

The attainable accuracy of the derived maser positions depends on the
size of the synthesized beam and the signal-to-noise ratio of the
detection, but the absolute positional accuracy of the
phase-referencing calibrator also has to be considered.  The VLA
calibrator manual classifies the position of J1755$-$2232 to be
between 0\farcs 002$-$0\farcs 01 accurate. Conservatively assuming the
larger value, an error of 10 mas would be dominating the error of the
derived maser positions.  By shifting the calibrator data during the
calibration procedure, we instead used the position obtained from the
Radio Fundamental Catalog (RFC\footnote{http://astrogeo.org/rfc/}),
providing sub-mas accuracies of 0.16 mas and 0.28 mas in R.A.\, and decl.,
respectively.  The position used was (J2000) R.A.\, 17$^h$55$^m$26.284535$^s$,
decl. $-$22$^\circ$32\arcmin10.61573\arcsec, which is 22 mas from the VLA
catalog position. As a result, the positional errors of the calibrator
combined with the VLA-derived errors are governed by the 0.8 and 2.4
mas VLA maser errors in R.A.\, and decl., respectively.  This is much
smaller than the typical quoted absolute values of 1\farcs 7, 0\farcs
035 and 0\farcs 060 errors, respectively, for the {\it MSX}, {\it WISE} and
2MASS catalogs, and of the same order as the error for the {\it Gaia}
catalog; the {\it Gaia} positional accuracy is thus directly
comparable to that of the derived VLA maser positions.

\subsection{Total and systematic offsets}
To determine how close the IR/optical catalog positions are to the VLA
SiO maser position, offsets were calculated between the VLA and {\it
  MSX}, 2MASS, {\it WISE}, and {\it Gaia} matches respectively.  Figure
\ref{offsets} plots the offset distribution, showing that the {\it
  MSX} positions can always be improved using any of the other
catalogs considered here. Furthermore,
positions are best matched using {\it Gaia} positions when available,
next followed by 2MASS. This is further illustrated in the scatter
plot of the positions as a function of R.A.\, and decl., where the {\it
  Gaia} and 2MASS positions are tightly clustered around the VLA maser
positions (Fig.\ \ref{radecscatter}).  The mean offsets
between the SiO masers and {\it MSX}, {\it WISE}, 2MASS and {\it Gaia}
catalogs are 0\farcs 89, 0\farcs 26, 0\farcs 12 and 0\farcs 01,
respectively. 

\begin{figure}[t]
\begin{center}
\includegraphics[width=8.5cm]{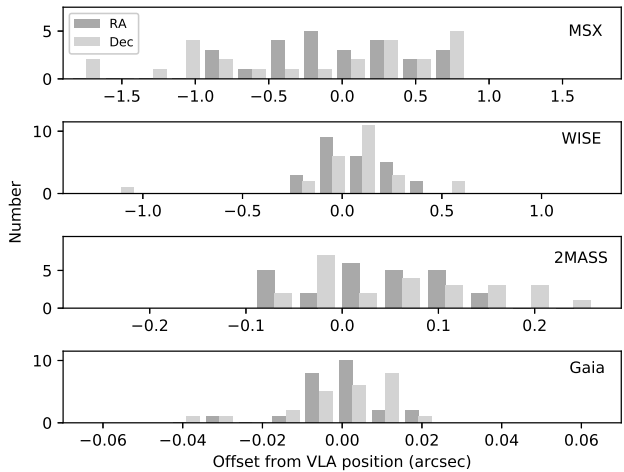}
\caption{R.A., (dark gray) and decl.\,(light gray) offsets in arcseconds between
  the VLA SiO maser positions and the {\it MSX}, {\it WISE}, 2MASS and {\it Gaia}
  positions. Note the difference in scale in the offsets for
  the different catalogs.\label{radecoff}}
\end{center}
\end{figure}

Along with Fig.\ \ref{radecscatter}, Fig.\ \ref{radecoff} separates
the spread in the positional offsets into R.A., and decl. components, in
order to consider any systematic offsets for any of the surveys.
There is a weak trend of the 2MASS data preferentially reporting a
more northern declination than the SiO masers, and conversely, that
the {\it MSX} data reports a more Southern declination.

\vspace*{0.5cm}
\section{Discussion}

The data confirm that {\it Gaia} positions are superior in pinpointing
the stellar SiO maser emission if it can be matched.  This is not
surprising, given that the SiO masers arise close to the central star,
at the inside of the larger circumstellar envelope (CSE) where dust
and other molecules are residing.  SiO masers are further known to be
ubiquitous in AGB stars with thin CSEs, allowing for the central star
to still be detectable in the optical.  If a {\it Gaia} match is not
obtainable, for example for more optically obscured or thicker shell
objects, 2MASS will provide a very good option for improving the
positional accuracy compared to any of the other IR catalogs.  We here
discuss reasons for why the {\it WISE} positions appear to be less accurate
in tracing the masers (\ref{wise2mass}) and how the detection rates in
the BAaDE SiO maser survey are improved using 2MASS positions
(\ref{baadedet}).

\subsection{{\it WISE} versus 2MASS positions}\label{wise2mass}
It is clear that the 2MASS positions are more accurate than {\it WISE} in
predicting the SiO maser positions, despite the better quoted
positional accuracy of {\it WISE}.  There are several possible causes for
this; first, {\it WISE} data has an intrinsically worse resolution (wider
point spread function).  By inspecting the fields around the targets
in the IRSA database, we noted that some of the 2MASS targets have
multiple possible matches, which likely are confused in the {\it WISE} data.
Secondly, the turnover of the SED for these sources tend to occur
around 1-2 $\mu$m, with the {\it WISE} bands at longer wavelengths being
sensitive for emission from the CSE. Like for {\it Gaia}, the shorter
2MASS wavelengths are more likely to directly probe the central star,
which will then better pinpoint the stellar position no matter how far
the CSE extends or how asymmetric the CSE is.  The SiO masers are
known to occur within a couple to a few stellar radii
\citep{diamond94,perrin15}, inside the main circumstellar shell and
close to the dust condensation radius, beyond which the SiO becomes
locked up in dust grains.  The {\it WISE} data will be more dominated by the
full extent of the CSE, and despite a quasi-spherical mass loss
assumed during the AGB phase, the larger size overall will likely make
it more difficult to exactly measure the position of the central
star. We assume that this effect is worse for {\it MSX}, which
operated at even longer wavelengths than {\it WISE}.

\subsection{Improved detection rate in the BAaDE survey}\label{baadedet}
As this program intended to improve initial positions from the {\it
  MSX} catalog and to assert what positions to use for VLBI, applying
{\it Gaia} or 2MASS positions to our VLA survey should improve our
maser detection rate.  For the 28,000 star BAaDE survey, we rely on an
assumed high detection rate initially and then utilize the detected
masers to perform self-calibration, applying the resulting phase
corrections to nearby targets lacking detections (L.O.\,Sjouwerman et
  al., in prep.).  The self-calibration
procedure prevents improved positional information to be derived, as
is usually obtained in the more commonly applied phase-referencing
scheme.  This strategy has the clear advantage that it removes
hundreds of calibration hours using the sparse high-frequency weather
needed for our survey, and still provides velocities along with
positions.  However, in regions of the sky where the source density is
lower our calibration scheme is less effective, and every detection is
crucial for the calibration of neighboring targets.  Especially for
weaker masers, a positional error of $1-2$\arcsec\ could result in the
maser not being detected in our pipeline which considers emission to
be at the phase center (read {\it MSX} position), thereby reducing the
effectiveness of our calibration strategy.  Exchanging the {\it MSX}
positions with {\it Gaia} or 2MASS positions in the self-calibration
scheme should improve our detection rate and thus the efficiency of
the survey overall.

\begin{figure}[t]
\begin{center}
\includegraphics[width=8cm]{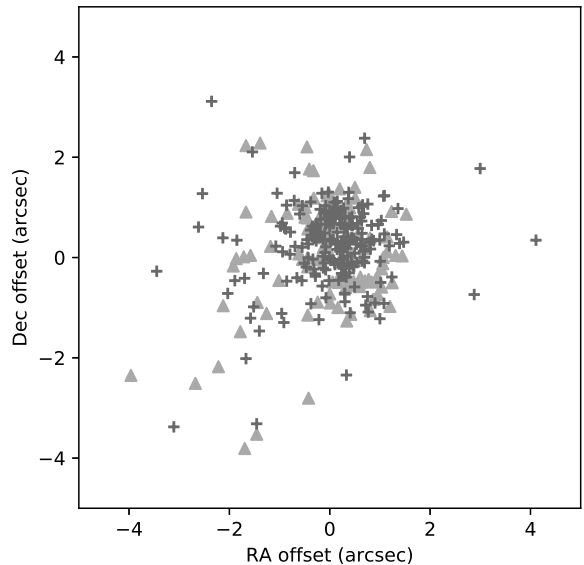}
\caption{Distribution of the shifts (in arcseconds) applied to a set
  of sources in a BAaDE observing run, from MSX to 2MASS
  positions. Triangles denote sources that were detected using the
  {\it MSX} positions, while the plus-signs indicate the
  \emph{additional detections} made by shifting targets to 2MASS
  positions.  No initial detections were lost in the
  process. \label{2masstest}}
\end{center}
\end{figure}

While approximately 30\% of our BAaDE sample may be expected to be
detected in the {\it Gaia} DR2 catalog (a full cross-match is
currently under way), 96\% have cross-matches in the 2MASS survey.
Hence we focused on using the 2MASS cross-matches in the VLA campaign.
While the {\it Gaia} positions would be preferable for VLBI 43 GHz
observing, the BAaDE survey is performed in C- and D-configuration at
the VLA.  With a resulting synthesized beamwidth of 0\farcs 5--
1\farcs 5, 2MASS positions are sufficiently accurate and there should
be little difference in the detection rate using {\it Gaia} instead of
2MASS positions.  We consequently tested our VLA BAaDE pipeline on a
random typical observing run by shifting the sources from the
originally observed {\it MSX} positions to their corresponding 2MASS
positions, and then re-running the pipeline.  While the original data
reduction reported 208 detections, using the 2MASS positions the
number increased to 347, thus a 40\% increase in the detection rate
(see Fig.\ \ref{2masstest}).  All of the originally detected sources
were detected with the 2MASS positions, thus the introduced shifts did
not shift any sources outside the beam.

\section{Conclusions}
By comparing stellar SiO maser positions derived from VLA
phase-referencing observations to those listed by the {\it MSX}, {\it WISE},
2MASS and {\it Gaia} catalogs, we have found that it is always
preferred to replace the {\it MSX} positions with positions from other
catalogs and that {\it Gaia} positions most closely match those of the
SiO masers (typically within $\sim$0\farcs 01).  For follow-up work, or new work
done by pre-selecting targets using IR colors, the results can be
significantly improved by performing cross-matching to either the
2MASS or the {\it Gaia} (matched to the 2MASS counterpart) catalogs
and using their positional information.  The mean offsets between the
SiO masers and {\it MSX}, {\it WISE}, 2MASS and {\it Gaia} catalogs are
0\farcs 89, 0\farcs 26, 0\farcs 12 and 0\farcs 01, respectively.  The
SiO maser emitting stars considered contain thin CSEs. For objects
with thicker shells, and for other work in optically obscured regions,
using 2MASS positions should be sufficient. For follow-up VLBI work,
additional matching to {\it Gaia} positions is clearly preferred.
\\

\acknowledgments 

The BAaDE project is funded by National Science Foundation Grants
1517970 (UNM) /1518271 (UCLA).

The National Radio Astronomy Observatory is a facility of the National
Science Foundation operated under cooperative agreement by Associated
Universities, Inc.

This research made use of data products from the
Midcourse Space Experiment. Processing of the data was funded by the
Ballistic Missile Defense Organization with additional support from
NASA Office of Space Science.

This research has also made use of the NASA/ IPAC Infrared Science Archive,
which is operated by the Jet Propulsion Laboratory, California
Institute of Technology, under contract with the National Aeronautics
and Space Administration.

This publication makes use of data products from the Two Micron All
Sky Survey, which is a joint project of the University of
Massachusetts and the Infrared Processing and Analysis
Center/California Institute of Technology, funded by the National
Aeronautics and Space Administration and the National Science
Foundation.

This publication makes use of data products from the Wide-field
Infrared Survey Explorer, which is a joint project of the University
of California, Los Angeles, and the Jet Propulsion
Laboratory/California Institute of Technology, funded by the National
Aeronautics and Space Administration.

This work has made use of data from the European Space Agency (ESA)
mission {\it Gaia} (\url{https://www.cosmos.esa.int/gaia}), processed
by the {\it Gaia} Data Processing and Analysis Consortium (DPAC,
\url{https://www.cosmos.esa.int/web/gaia/dpac/consortium}). Funding
for the DPAC has been provided by national institutions, in particular
the institutions participating in the {\it Gaia} Multilateral
Agreement. 

\vspace{5mm} \facilities{VLA, IRSA, WISE, MSX, 2MASS,
  Gaia}


\begin{thebibliography}{}
\bibitem[Belokurov et al.(2017)]{belokurov17}Belokurov, V., Erkal,
  D., Deason, A.J., Koposov, S.E., De Angeli, F.,
  Evans, D.W., Fraternali, F., Mackey, D.\ 2017, \mnras, 446, 4711
\bibitem[Diamond et al.(1994)]{diamond94}Diamond, P.J., Kemball, A.J.,
  Junor, W., et al.\ 1994, \apj, 430, L61
\bibitem[Egan et al.(2003)]{egan03}Egan, M.P., Price, S.D., Kraemer,
  K.E., et al.\ 2003, ‘The Midcourse Space Experiment Point Source
  Catalog, Version 2.3’, Air Force Research Laboratory Technical
  Report AFRL-VS-TR-2003-1589 (Springfield, VA: NTIS)
\bibitem[Evans et al.(2018)]{evans18}Evans, D.W., Riello, M., De Angeli, F. et
  al., 2018, \aap, 614, A4
\bibitem[Gaia Collaboration et al.(2016)]{gaiacoll16}Prusti, T., de
  Bruijne, J.H.J., Brown, A., et al., 2016, \aap, 595, A1
\bibitem[Gaia Collaboration et al.(2018)]{gaiacoll18}Brown, A.G.A.,
  Vallenari, A., et al., 2018, \aap, 616, A1
\bibitem[Habing et al.(1996), and references therein]{habing96}Habing,
  H.\ 1996, \araa, 7, 97
\bibitem[Lindegren et al.(2018)]{lindegren18}
Lindegren, L., Hern\'andez, J., Bombrun, ,A., et al., 2018, A\&A, 616, A2
\bibitem[Perrin et al.(2015)]{perrin15}Perrin, G., Cotton, W.D.,
  Millan-Gabet, R., et al.\ 2015, \aap, 576 70
\bibitem[Quiroga-Nu{\~n}ez et al.(2018)]{quiroga18} Quiroga-Nu{\~n}ez,
  L.H., van Langevelde, H.J., Pihlstr{\"o}m, Y.M., et al., 2018, in:
  Recio-Blanca, A., de Laverny, P., Brown, A.G.A., \& Prusti, T.,
  (eds.) , Astrometry and Astrophysics in the Gaia sky, Proc. IAU
  Symposium No. 330, Volume 330, p. 245
\bibitem[Reid et al.(1988)]{reid88}Reid, M.J., Schneps, M.H., Moran,
  J.M., et al.\ 1988, \apj, 330, 809
\bibitem[Sjouwerman et al.(2009)]{sjouwerman09}Sjouwerman, L.O.,
  Capen, S.M., \& Claussen, M.J., 2009, \apj, 705, 1554
\bibitem[Skrutskie et al.(2006)]{skruts06}Skrutskie, M.F., Cutri, R.M.,
  Weinberg, R., et al., 2006, \aj, 131, 1163
\bibitem[Trapp et al.(2018)]{trapp18}Trapp, A.C., Rich, R.M., Morris,
  M.R., Sjouwerman, et al.\ 2018, \apj, 861, 75
\bibitem[Wright et al.(2010)]{wright10}Wright, E.L., Eisenhardt, P.R.M.,
  Mainzer, A.K., et al., 2010, \aj, 140, 1868
\end{thebibliography}
\end{document}